\documentclass[aps,prl,twocolumn,superscriptaddress,nofootinbib]{revtex4}

\usepackage[dvipsnames]{xcolor}
\usepackage{empheq}
\usepackage{gensymb}
\usepackage{graphicx}
\usepackage{dcolumn}
\usepackage{bm}
\usepackage{hyperref}
\usepackage[mathlines]{lineno}

\hypersetup{
    pdfnewwindow=true,      
    colorlinks=true,       
    linkcolor=blue,          
    citecolor=blue,        
    filecolor=blue,      
    urlcolor=blue           
}

\newcommand{\nm}{\nu_\mu}
\newcommand{\enu}{E_{\nu}}
\newcommand{\difd}{\textrm{d}}
\newcommand{\dEdx}{\difd E/\difd x}
\newcommand{\proton}{\textrm{p}}
\newcommand{\pproton}{p_\proton}
\newcommand{\thetaproton}{\theta_\proton}
\newcommand{\pmu}{p_\mu}
\newcommand{\thetamu}{\theta_\mu}
\newcommand{\neutron}{\textrm{n}}
\newcommand{\mneutron}{m_\neutron}
\newcommand{\pn}{p_\neutron}

\newcommand{\pimtv}{\vec{p}_\textrm{IMT}}

\newcommand{\pfmv}{\vec{p}_\textrm{FM}}
\newcommand{\dalphat}{\delta\alpha_\textrm{T}}
\newcommand{\pt}{p_\textrm{T}}
\newcommand{\dpt}{\delta \pt}
\newcommand{\dptv}{\delta\vec{p}_\textrm{T}}
\newcommand{\dplv}{\delta p_\textrm{L}}
\newcommand{\dpv}{\delta\vec{p}}
\newcommand{\dpm}{\delta p}
\newcommand{\gevc}{\textrm{GeV}/c}
\newcommand{\gevcc}{\textrm{GeV}/c^2}

\newcommand{\nucleus}{\textrm{A}}
\newcommand{\ptl}{\vec{p}_\textrm{T}^{\,\mu}}
\newcommand{\ptlm}{p_\textrm{T}^{\,\mu}}
\newcommand{\plong}{p_\textrm{L}}
\newcommand{\pll}{\plong^{\,\mu}}
\newcommand{\plp}{\plong^{\,\proton}}
\newcommand{\dpl}{\delta\plong}
\newcommand{\ptni}{\vec{p}_\textrm{T}^{\,\proton}}
\newcommand{\maqe}{M^\textrm{QE}_\textrm{A}}

\newcommand{\aprime}{\nucleus^\prime}
\newcommand{\fshadron}{\textrm{X}}
\newcommand{\amass}{m_\nucleus}
\newcommand{\apmass}{m_{\aprime}}
\newcommand{\genie}{\textsc{genie}}
\newcommand{\gver}{2.8.4}
\newcommand{\nuwro}{\textsc{n}u\textsc{w}ro}
\newcommand{\pythia}{\textsc{pythia6}}
\newcommand{\geant}{\textsc{geant4}}

\newcommand{\mnvgenie}{\textsc{m}nv\genie-v1}
\newcommand{\nofsi}{\emph{no-FSI}}
\newcommand{\lside}{\emph{left}}
\newcommand{\rside}{\emph{right}}

\let\oldalign\align
\let\oldendalign\endalign

\renewenvironment{align}
  {\linenomathNonumbers\oldalign}
  {\oldendalign\endlinenomath}

\let\oldempheq\empheq
\let\oldendempheq\endempheq

\renewenvironment{empheq}
  {\linenomathNonumbers\oldempheq}
  {\oldendempheq\endlinenomath}

\begin{document}

\title{
Measurement of final-state correlations in neutrino muon-proton mesonless production on hydrocarbon at $\langle E_\nu\rangle=3$ GeV \\
}

\newcommand{\Rutgers}{Rutgers, The State University of New Jersey, Piscataway, New Jersey 08854, USA}
\newcommand{\Hampton}{Hampton University, Dept. of Physics, Hampton, VA 23668, USA}
\newcommand{\Dortmund}{Institute of Physics, Dortmund University, 44221, Germany }
\newcommand{\Otterbein}{Department of Physics, Otterbein University, 1 South Grove Street, Westerville, OH, 43081 USA}
\newcommand{\JMU}{James Madison University, Harrisonburg, Virginia 22807, USA}
\newcommand{\Florida}{University of Florida, Department of Physics, Gainesville, FL 32611}
\newcommand{\UCIrvine}{Department of Physics and Astronomy, University of California, Irvine, Irvine, California 92697-4575, USA}
\newcommand{\CBPF}{Centro Brasileiro de Pesquisas F\'{i}sicas, Rua Dr. Xavier Sigaud 150, Urca, Rio de Janeiro, Rio de Janeiro, 22290-180, Brazil}
\newcommand{\PUCP}{Secci\'{o}n F\'{i}sica, Departamento de Ciencias, Pontificia Universidad Cat\'{o}lica del Per\'{u}, Apartado 1761, Lima, Per\'{u}}
\newcommand{\INRM}{Institute for Nuclear Research of the Russian Academy of Sciences, 117312 Moscow, Russia}
\newcommand{\Jlab}{Jefferson Lab, 12000 Jefferson Avenue, Newport News, VA 23606, USA}
\newcommand{\Pittsburgh}{Department of Physics and Astronomy, University of Pittsburgh, Pittsburgh, Pennsylvania 15260, USA}
\newcommand{\Guanajuato}{Campus Le\'{o}n y Campus Guanajuato, Universidad de Guanajuato, Lascurain de Retana No. 5, Colonia Centro, Guanajuato 36000, Guanajuato M\'{e}xico.}
\newcommand{\Athens}{Department of Physics, University of Athens, GR-15771 Athens, Greece}
\newcommand{\Tufts}{Physics Department, Tufts University, Medford, Massachusetts 02155, USA}
\newcommand{\WM}{Department of Physics, College of William \& Mary, Williamsburg, Virginia 23187, USA}
\newcommand{\FNAL}{Fermi National Accelerator Laboratory, Batavia, Illinois 60510, USA}
\newcommand{\Purdue}{Department of Chemistry and Physics, Purdue University Calumet, Hammond, Indiana 46323, USA}
\newcommand{\MCLA}{Massachusetts College of Liberal Arts, 375 Church Street, North Adams, MA 01247}
\newcommand{\UMD}{Department of Physics, University of Minnesota -- Duluth, Duluth, Minnesota 55812, USA}
\newcommand{\Northwestern}{Northwestern University, Evanston, Illinois 60208}
\newcommand{\UNI}{Universidad Nacional de Ingenier\'{i}a, Apartado 31139, Lima, Per\'{u}}
\newcommand{\Rochester}{University of Rochester, Rochester, New York 14627 USA}
\newcommand{\Austin}{Department of Physics, University of Texas, 1 University Station, Austin, Texas 78712, USA}
\newcommand{\USM}{Departamento de F\'{i}sica, Universidad T\'{e}cnica Federico Santa Mar\'{i}a, Avenida Espa\~{n}a 1680 Casilla 110-V, Valpara\'{i}so, Chile}
\newcommand{\Geneva}{University of Geneva, 1211 Geneva 4, Switzerland}
\newcommand{\Chicago}{Enrico Fermi Institute, University of Chicago, Chicago, IL 60637 USA}
\newcommand{\hired}{}
\newcommand{\OregonState}{Department of Physics, Oregon State University, Corvallis, Oregon 97331, USA}
\newcommand{\oxford}{Oxford University, Department of Physics, Oxford, United Kingdom}
\newcommand{\umiss}{University of Mississippi, Oxford, Mississippi 38677, USA}
\newcommand{\upenn}{Department of Physics and Astronomy, University of Pennsylvania, Philadelphia, PA 19104}
\newcommand{\AMU}{AMU Campus, Aligarh, Uttar Pradesh 202001, India}
\newcommand{\wroclaw}{University of Wroclaw, plac Uniwersytecki 1, 50-137 Wrocław, Poland}
\newcommand{\Mohali}{Knowledge city, Sector 81, SAS Nagar, Manauli PO 140306}
\newcommand{\chrismarshallThanks}{now at Lawrence Berkeley National Laboratory, Berkeley, CA 94720, USA}
\newcommand{\joelmousseauThanks}{now at University of Michigan, Ann Arbor, MI 48109, USA}
\newcommand{\cpatrickThanks}{Now at University College London, London WC1E 6BT, UK}
\newcommand{\jwolcottThanks}{now at Tufts University, Medford, MA 02155, USA}

\author{X.-G.~Lu}                         \affiliation{\oxford}
\author{M.~Betancourt}                    \affiliation{\FNAL}
\author{T.~Walton}            \affiliation{\FNAL}

\author{F.~Akbar}                         \affiliation{\AMU}
\author{L.~Aliaga}                        \affiliation{\WM}  \affiliation{\PUCP}
\author{O.~Altinok}                       \affiliation{\Tufts}
\author{D.A.~Andrade}                     \affiliation{\Guanajuato}
\author{M.~Ascencio}                      \affiliation{\PUCP}
\author{L.~Bellantoni}                    \affiliation{\FNAL}
\author{A.~Bercellie}                     \affiliation{\Rochester}
\author{A.~Bodek}                         \affiliation{\Rochester}
\author{A.~Bravar}                        \affiliation{\Geneva}
\author{H.~Budd}                          \affiliation{\Rochester}
\author{T.~Cai}                           \affiliation{\Rochester}
\author{M.F.~Carneiro}                    \affiliation{\OregonState}
\author{J.~Chaves}                        \affiliation{\upenn}
\author{D.~Coplowe}                       \affiliation{\oxford}
\author{H.~da~Motta}                      \affiliation{\CBPF}
\author{S.A.~Dytman}                      \affiliation{\Pittsburgh}
\author{G.A.~D\'{i}az~}                   \affiliation{\Rochester}  \affiliation{\PUCP}
\author{J.~Felix}                         \affiliation{\Guanajuato}
\author{L.~Fields}                        \affiliation{\FNAL}  \affiliation{\Northwestern}
\author{R.~Fine}                          \affiliation{\Rochester}
\author{A.M.~Gago}                        \affiliation{\PUCP}
\author{R.Galindo}                        \affiliation{\USM}
\author{H.~Gallagher}                     \affiliation{\Tufts}
\author{A.~Ghosh}                         \affiliation{\USM}  \affiliation{\CBPF}
\author{R.~Gran}                          \affiliation{\UMD}
\author{D.A.~Harris}                      \affiliation{\FNAL}
\author{S.~Henry}                         \affiliation{\Rochester}
\author{S.~Jena}                          \affiliation{\Mohali}
\author{D.Jena}                           \affiliation{\FNAL}
\author{J.~Kleykamp}                      \affiliation{\Rochester}
\author{M.~Kordosky}                      \affiliation{\WM}
\author{T.~Le}                            \affiliation{\Tufts}  \affiliation{\Rutgers}
\author{E.~Maher}                         \affiliation{\MCLA}
\author{S.~Manly}                         \affiliation{\Rochester}
\author{W.A.~Mann}                        \affiliation{\Tufts}
\author{C.M.~Marshall}\thanks{\chrismarshallThanks}  \affiliation{\Rochester}
\author{K.S.~McFarland}                   \affiliation{\Rochester}  \affiliation{\FNAL}
\author{A.M.~McGowan}                     \affiliation{\Rochester}
\author{B.~Messerly}                      \affiliation{\Pittsburgh}
\author{J.~Miller}                        \affiliation{\USM}
\author{A.~Mislivec}                      \affiliation{\Rochester}
\author{J.G.~Morf\'{i}n}                  \affiliation{\FNAL}
\author{J.~Mousseau}\thanks{\joelmousseauThanks}  \affiliation{\Florida}
\author{D.~Naples}                        \affiliation{\Pittsburgh}
\author{J.K.~Nelson}                      \affiliation{\WM}
\author{C.~Nguyen~}                       \affiliation{\Florida}
\author{A.~Norrick}                       \affiliation{\WM}
\author{Nuruzzaman}                       \affiliation{\Rutgers}  \affiliation{\USM}
\author{A.~Olivier}                       \affiliation{\Rochester}
\author{V.~Paolone}                       \affiliation{\Pittsburgh}
\author{C.E.~Patrick}\thanks{\cpatrickThanks}  \affiliation{\Northwestern}
\author{G.N.~Perdue}                      \affiliation{\FNAL}  \affiliation{\Rochester}
\author{M.A.~Ram\'{i}rez}                 \affiliation{\Guanajuato}
\author{R.D.~Ransome}                     \affiliation{\Rutgers}
\author{L.~Ren}                           \affiliation{\Pittsburgh}
\author{D.~Rimal}                         \affiliation{\Florida}
\author{P.A.~Rodrigues}                   \affiliation{\umiss}  \affiliation{\Rochester}
\author{D.~Ruterbories}                   \affiliation{\Rochester}
\author{H.~Schellman}                     \affiliation{\OregonState}  \affiliation{\Northwestern}
\author{J.T.~Sobczyk}                     \affiliation{\wroclaw}
\author{C.J.~Solano~Salinas}              \affiliation{\UNI}
\author{H.~Su}                            \affiliation{\Pittsburgh}
\author{M.~Sultana}                       \affiliation{\Rochester}
\author{E.~Valencia}                      \affiliation{\WM}  \affiliation{\Guanajuato}
\author{D.~Wark}                          \affiliation{\oxford}
\author{A.~Weber}                         \affiliation{\oxford}
\author{J.~Wolcott}\thanks{\jwolcottThanks}  \affiliation{\Rochester}
\author{M.Wospakrik}                      \affiliation{\Florida}
\author{B.~Yaeggy}                        \affiliation{\USM}


\collaboration{The MINERvA Collaboration}\ \noaffiliation
\date{\today}

\begin{abstract}

Final-state kinematic imbalances are measured in mesonless production of $\nm + \nucleus \to \mu^- + \proton + \fshadron$ in the MINERvA tracker. Initial- and final-state nuclear effects are probed using the direction of the $\mu^-$-$\proton$ transverse momentum imbalance and the initial-state momentum of the struck neutron.
Differential cross sections are compared to predictions based on current approaches to medium modeling.
These models under-predict the cross section 
at intermediate intranuclear momentum transfers that generally exceed the Fermi momenta.
 As neutrino interaction models need to correctly incorporate the effect of the nucleus in order to predict neutrino energy resolution in oscillation experiments, this result points to a region of phase space where additional cross section strength is needed in current models, and demonstrates a new technique that would be suitable for use in fine grained liquid argon detectors where the effect of the nucleus may be even larger.  

\end{abstract}

\maketitle

An accurate understanding of nuclear medium modifications to neutrino-nucleon interactions is required for reliable measurements of fundamental neutrino properties.
The distributions of final-state observables reflect complicated and intertwined effects from nucleon and  nuclear dynamics, and the interpretation of single-particle kinematics is thereby obscured~\cite{Nieves:2011yp}.  These underlying dynamics can influence  neutrino energy reconstruction in oscillation experiments~\cite{Coloma:2013rqa, Abe:2017vif}.
Certain categories of nuclear effects, however, can  be separated by variables~\cite{Lu:2015tcr,Furmanski:2016wqo} designed to elicit final-state correlations that are absent for neutrino interactions on free nucleons, but are in-play in neutrino-nucleus scattering. 
This Letter reports the  measurements with such variables for the purpose of constraining nuclear effects in neutrino interactions.

In charged-current neutrino-nucleus scattering there is an imbalance, $\dpv$, between the initial neutrino momentum and the sum of final-state lepton and hadron momenta as a result of nuclear effects.  This imbalance is the sum of Fermi motion (FM) and intranuclear momentum transfer (IMT), which is the sum of all other effects including nucleon correlations~\cite{Singh:1992dc, Gil:1997bm, Nieves:2004wx, Nieves:2005rq, Martini:2009uj, Benhar:1994hw,  Delorme:1985ps, Marteau:1999jp, Martini:2010ex, Nieves:2011pp, Nieves:2011yp, Martini:2011wp} and final-state interactions (FSI):
\begin{align}
\dpv=\pfmv-\pimtv.\label{eq:dpfmimt}
\end{align}

In $\nu_{\mu}$ charged-current quasielastic (QE) interactions, $\nm + \neutron \to \mu^- + \proton$, 
momentum is transferred from the leptonic current to the target neutron. However if the neutron is correlated with other nucleons, the momentum transfer is shared among the correlated partners. 
In both cases, final-state interactions occur as particles from the primary interaction propagate through the nucleus exchanging energy, momentum and charge with the nuclear environment.  Primary particles can be absorbed during this propagation; baryonic resonances (RES) can be produced in the primary interaction and the resulting products can undergo FSI and be absorbed.  These non-QE processes give the same observable final state as QE scattering. 

\begin{figure}[!ht]
\centering
\includegraphics[width=0.6\columnwidth]{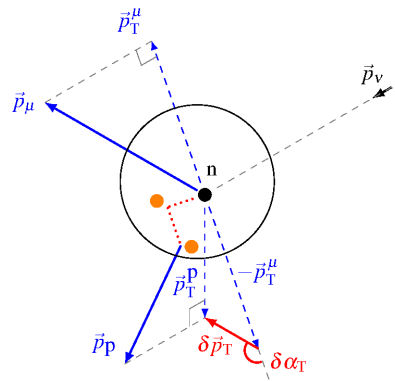}
\caption{Schematic definition of the transverse kinematics~\cite{Lu:2015tcr}.}\label{fig:transdia}
\end{figure}

This study focuses on the QE-like process
\begin{align}
\nm+\nucleus\to\mu^-+\proton+\fshadron,\label{eq:sigchannel}
\end{align}
where $\fshadron$ is a final-state hadronic system consisting of the nuclear remnant with possible additional protons but without pions that indicate RES or other processes. 
In Eq.~(\ref{eq:sigchannel}), the incident neutrino energy, $\enu$, is unknown, but the dependence of $\dpv$ on $\enu$ can be removed. This can be done as follows:

Firstly, decompose $\dpv$  into longitudinal and transverse components with respect to the  neutrino direction, 
\begin{empheq}[left=\empheqlbrace]{align}
\dpv&\equiv(\dplv,~\dptv),\\
\enu&= \pll + \plp -\dplv \label{eq:dpl},\\
\vec{0}&= \ptl + \ptni -\dptv \label{eq:dpt}, 
\end{empheq}
where $\vec{p}^{\,\mu}$ and $\vec{p}^{\,\proton}$ are the muon and proton momenta, respectively. The direction of the transverse momentum imbalance $\dptv$ (see schematic definition in Fig.~\ref{fig:transdia}),
\begin{align}
\dalphat&\equiv\arccos\frac{-\ptl\cdot\dptv}{\ptlm\dpt},\label{eq:datdef}
\end{align}
is uniformly distributed in the absence of IMT because of the isotropic nature of Fermi motion.  This variable is thus sensitive to IMT~\cite{Lu:2015tcr}.
Because $\left|\ptni\right|>\left|\ptl\right|$ for $\dalphat<90^\circ$, accelerating FSI can be distinguished from decelerating FSI using $\dalphat$. 
Recent measurements of $\dpt$ and $\dalphat$  on hydrocarbon at beam energy around 600~MeV by the T2K Collaboration can be found in Ref.~\cite{Abe:2018pwo}. 

Secondly, under the  assumption that $\fshadron$ is just the  remnant nucleus, $\nucleus^{\prime}$, then $\dpm$ gives the magnitude of its recoil momentum, and
\begin{empheq}[left=\empheqlbrace]{align}
\dpl&=\frac{1}{2}R-\frac{\apmass^2+\dpt^2}{2R},\label{eq:dplsolution}\\
R&\equiv\amass+\pll+\plp-E^\mu-E^\proton,\label{eq:rr}
\end{empheq}
where $m_{\nucleus^{(\prime)}}$, $E^{\mu\,(\proton)}$  are the nuclear target (remnant) mass, and the muon (proton) energy, respectively~\cite{Furmanski:2016wqo}. 
In the limit of zero IMT (that is, pure QE), the recoiling momentum of $\aprime$ balances the initial neutron momentum and
\begin{align}
\pn&=\dpm\label{eq:pn},
\end{align}
which can be estimated using the relation $\apmass=\amass-\mneutron+b$, where $\mneutron$ is the neutron mass, and $b=+27.13~\textrm{MeV}$ for carbon obtained from the probabilistic model for excitation energy~\cite{Furmanski:2016wqo}.

This Letter presents the measurement of $\dalphat$ and $\pn$ in $\nu_\mu$ induced production on polystyrene by the MINERvA experiment.  The signal is defined as an event with no pions, one muon and at least one proton satisfying
\begin{empheq}[left=\empheqlbrace]{align}
1.5~\gevc<&\pmu<10~\gevc,~\thetamu<20^\circ,\label{eq:pmuon}\\
0.45~\gevc<&\pproton<1.2~\gevc,~\thetaproton<70^\circ,\label{eq:thetaproton}
\end{empheq}
where $\pmu$ and $\thetamu$ ($\pproton$ and $\thetaproton$) are the muon (proton) momentum and polar angle with respect to the neutrino direction, respectively, when exiting the nucleus.  Nuclear effects in terms of Fermi motion and IMT are measured and compared to model predictions. 

The MINERvA experiment is in the NuMI beam line~\cite{Adamson:2015dkw} at Fermilab.  The detector is described in detail elsewhere~\cite{Aliaga:2013uqz}.  The tracker is constructed of hexagonal planes which are approximately perpendicular to the incoming neutrino beam and made from triangular scintillator strips.  Scintillator strips in adjacent planes are rotated by 60$^\circ$  with respect to each other, permitting three-dimensional track reconstruction which is efficient up to 70$^\circ$ from the detector axis. The scintillator is embedded in polystyrene, containing the carbon target nuclei. 
 The MINOS Near Detector is two meters downstream of the MINERvA detector and serves as a magnetized muon spectrometer~\cite{Michael:2008bc}.  The data used in this analysis corresponds to 3.28$\times10^{20}$ protons on target (POT) delivered from 2010 to 2012; the integrated $\nu_\mu$ flux prediction  ($2.88\times10^{-8}/\textrm{cm}^2/\textrm{POT}$) is from~\cite{Aliaga:2016oaz}.

Neutrino interactions are simulated with \genie~\gver~\cite{Andreopoulos:2009rq} in both a nominal form, with and without FSI, and also with a MINERvA `tune', (\mnvgenie).  The nuclear initial state is modeled as relativistic Fermi gas~\cite{Bodek:1980ar}.  Quasielastic~\cite{LlewellynSmith:1971uhs}, RES, and deep inelastic scattering (DIS) kinematics are modeled with a dipole axial form factor using $\maqe=0.99~\gevcc$, the Rein-Sehgal model~\cite{Rein:1980wg}, and the 2003 Bodek-Yang model~\cite{Bodek:2002ps}, respectively. \pythia~\cite{Sjostrand:2006za} and  models based on Koba-Nielsen-Olesen scaling~\cite{Koba:1972ng} are used to describe hadronization.  The $hA$ option of \genie~was used to model FSI~\cite{Dytman:2011zz}. The general performance of the \genie~FSI treatment for pions has been demonstrated in data versus simulation comparisons published by MINERvA~\cite{Eberly:2014mra, Aliaga:2015wva,McGivern:2016bwh, Altinok:2017xua, Betancourt:2017uso}. 

\mnvgenie~includes two-particle-two-hole (2p2h) excitations of the nucleus as formulated in the Valencia model~\cite{Nieves:2011yp, Sobczyk:2012ms, Gran:2013kda, Schwehr:2016pvn}.  The interaction strength with 2p2h has been tuned to MINERvA inclusive scattering data~\cite{Rodrigues:2015hik}, resulting in a significant enhancement relative to the Valencia model in a restricted region of energy-momentum transfer.  \mnvgenie~also includes a modification to the nonresonant pion production as constrained by deuterium data~\cite{Wilkinson:2014yfa}, and collective excitations of the nucleus for the QE channel. The latter are approximated as a superposition of one-particle-one-hole (1p1h) excitations and calculated with the Random Phase Approximation~\cite{Nieves:2004wx}. Because the affected events contribute little to the sample, the effects of nonresonant pion production and RPA in this analysis are negligible. 

In the nominal \genie~configuration, FSI are further categorized as follows:
\begin{enumerate}
\item{\it Noninteracting proton FSI} - FSI without pion absorption in which the proton propagates as a free particle.
\item{\it Accelerating proton FSI} - FSI without pion absorption in which the proton energy increases as a result of FSI.
\item{\it Decelerating proton FSI} - FSI without pion absorption in which the proton energy decreases as a result of FSI.
\item{\it Pion FSI} - FSI in which a pion is absorbed.
\end{enumerate}

In addition to \genie, data is also compared to the predictions of \nuwro~\cite{Golan:2012wx}. The initial state is modeled either as a local Fermi gas or with a Spectral Function~\cite{Benhar:1994hw}.  FSI are treated as  intranuclear cascades of hadronic interactions~\cite{Pandharipande:1992zz} incorporating the Oset model~\cite{Salcedo:1987md}, and 2p2h excitations are from the Valencia model~\cite{Nieves:2011yp, Sobczyk:2012ms}.

Events with at least two reconstructed tracks in the MINERvA tracker satisfying Eqs.~(\ref{eq:pmuon})-(\ref{eq:thetaproton}) are selected. The muon candidate track must match a track in the MINOS Near Detector, necessitating Eq.~(\ref{eq:pmuon}).  The two tracks are combined to determine $\pmu$ and $\thetamu$, with resolutions of $\sim$8\% at 5 $\gevc$ and $\sim$0.6 degrees, respectively.
The proton candidate is distinguished from possible mesons with specific energy loss ($\dEdx$). Its momentum $\pproton$ is determined according to the $\dEdx$ profile along the track~\cite{Walton:2014dka, Walton:2014esl}.  
If two or more protons satisfy Eq.~(\ref{eq:thetaproton}) (less than 1\% of all  selected events), the most energetic one    is taken as the proton candidate.

To improve $\pproton$ resolution, additional selection requirements to obtain elastically scattered and contained (ESC) protons~\cite{Lu:2016mjf} are introduced.  When a proton is not contained in the tracker or undergoes inelastic scattering it has a deteriorated momentum estimate.
ESC protons are selected by requiring large $\dEdx$  near the track endpoints. This reduces the spread of the reconstructed $\pproton$ to about 60\% of its previous measurement~\cite{Walton:2014esl}, resulting in a resolution of $\sim$2\% at 1~$\gevc$, at the cost of a reduction of statistics to about 40\% of the initial proton sample. 

\begin{figure}[!ht]
\centering
\includegraphics[width=0.99\columnwidth]{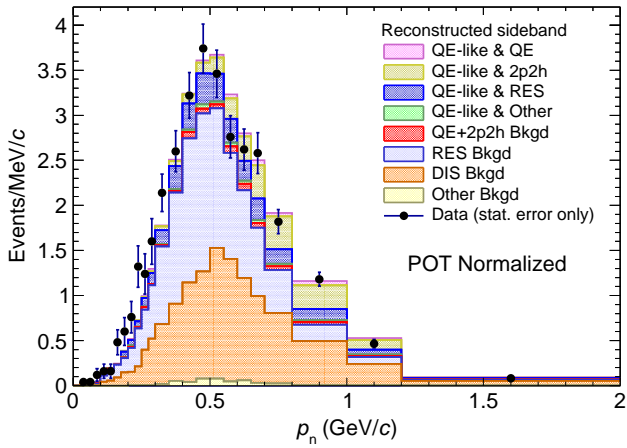} 
\caption{Reconstructed data distribution of $\pn$ in a sideband region, compared to full \mnvgenie+\geant~simulation after background tuning.  
The QE-like signal and the background are  decomposed into interaction modes.}\label{fig:sideband}
\end{figure}

The  efficiency (including acceptance effects) of event selection is estimated to be  8.6\% and the purity is 78\%.  The predicted  background contributions mainly come from RES where the pion from baryon decay exits the nucleus but is not identified (13.4\%) and DIS (5.4\%).  A data-driven method~\cite{Walton:2014esl} is used to determine backgrounds.  Sidebands are determined in the plane of unattached visible energy away from the interaction vertex {\it vs.} transferred four-momentum squared, $Q^2$.  Backgrounds from the \genie~simulation are rescaled so that the sidebands describe the data (See Fig.~\ref{fig:sideband}) and are then extrapolated into the signal region.  The background-subtracted distributions are then unfolded~\cite{DAgostini:1994fjx} with four iterations, where the number of iterations is chosen to balance between the bias and fluctuation of the unfolded distributions.  After a subsequent efficiency correction, event distributions are normalized by the product of the number of target nucleons ($3.11\times10^{30}$), POT, and $\nu_\mu$ flux, to obtain the flux-averaged differential cross sections. 

Systematic uncertainties are evaluated for all observables.  As an example, the  cross section uncertainty in $\dalphat$ is summarized here.  Besides statistical uncertainty (5-7\%), uncertainties arising from the NuMI flux  prediction (6\%), \genie~modeling (6-9\%) and detector  response (6-16\%) are accounted for.  For the latter two, parameters in the physics and detector models are varied within uncertainties and the resulting cross section variations are the assigned systematic uncertainties~\cite{Walton:2014esl}.  \genie~model uncertainties predominantly arise from modeling 2p2h, while the transverse projection of the muon and proton kinematics and the ESC proton selection have significant contributions from detector systematics. The total uncertainty is  20\% at $\dalphat=0^\circ$ and 180$^\circ$, and decreases to about 13\% at 90$^\circ$.

\begin{figure}[!ht]
\centering
\includegraphics[width=0.99\columnwidth]{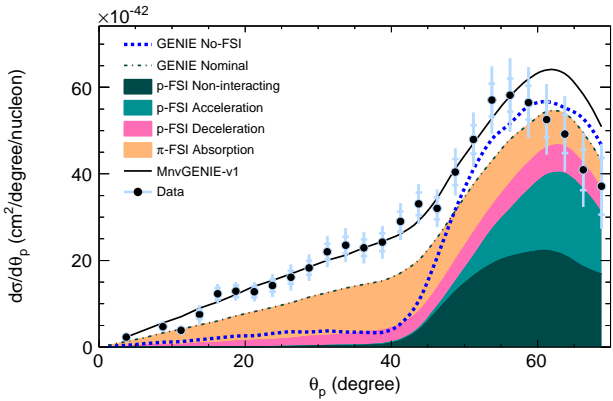}
\caption{Differential cross section in   $\thetaproton$  compared to \genie~predictions (nominal \genie~version: \gver).  Outer error bars represent the combined statistical and systematic uncertainties, while the inner ones statistical only.
 The white space between the lines for nominal \genie~and \mnvgenie~is mostly from the tuned 2p2h component in the latter. 
}\label{fig:protontheta}
\end{figure}

\begin{figure}[!ht]
\centering
\includegraphics[width=0.99\columnwidth]{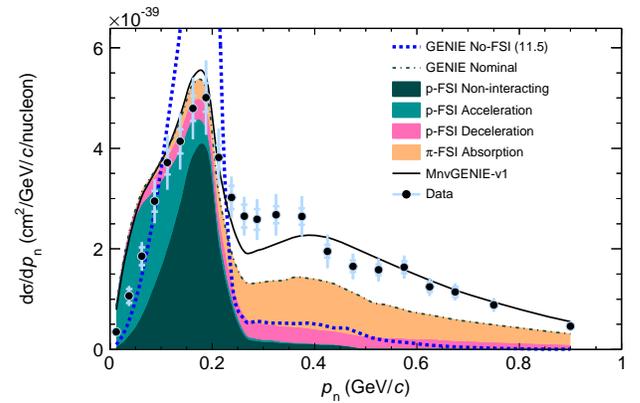}
\caption{Differential cross sections in  $\pn$, compared to \genie~predictions. The  \nofsi~prediction maximum is $11.5$ (see legend) in unit of $10^{-39}$ cm$^2/\gevc/\textrm{nucleon}$. 
}\label{fig:pn}
\end{figure}

\begin{figure}[!ht]
\centering
\includegraphics[width=0.99\columnwidth]{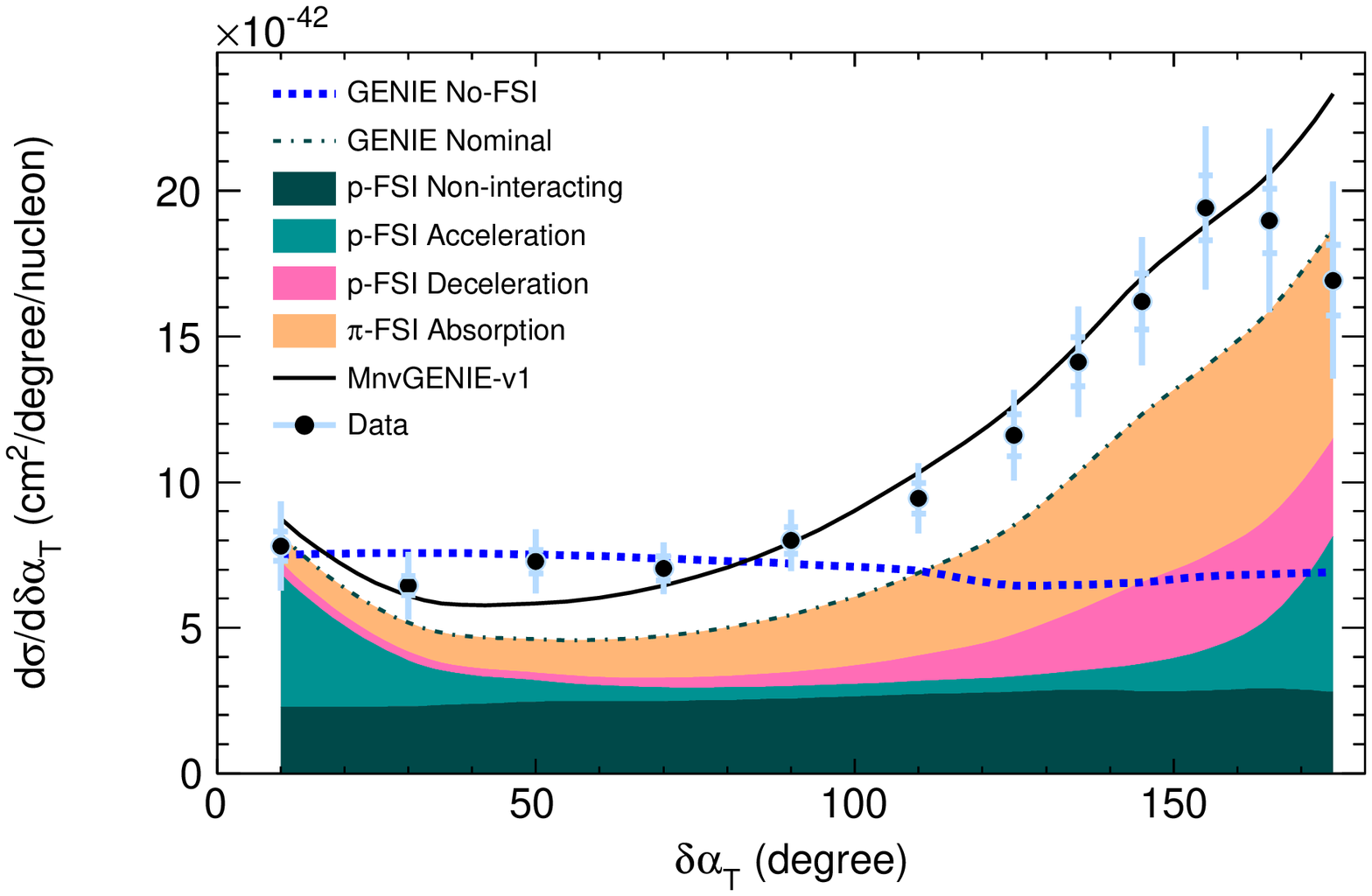}
\caption{Differential cross sections in  $\dalphat$, compared to \genie~predictions.  The peak in accelerating proton FSI on the right side is from events where  
$\ptni$~has a large angular separation from $-\ptl$ despite 
$\left|\ptni\right|>\left|\ptl\right|$.
}\label{fig:dat}
\end{figure}

Muon momentum and polar angle distributions are satisfactorily described by \mnvgenie~(Supplemental Material~1 Fig.~1), as is the proton momentum distribution (Supplemental Material~1 Fig.~2).  
There is a significant  over-prediction beyond 1-$\sigma$ total uncertainty at  
$\thetaproton\gtrsim 60^\circ$ (Fig.~\ref{fig:protontheta}).  Of the four FSI categories, noninteracting proton and accelerating proton FSI
are the ones that appear only in this high angle region.  Away from it, at $\thetaproton$ below $40^\circ$ where pion FSI and 2p2h contributions are large, the \mnvgenie~prediction describes the data very well.

For \genie~without FSI and nominal \genie~with noninteracting proton, the pure QE assumption of Eqs.~(\ref{eq:dplsolution})-(\ref{eq:pn}) is valid. The corresponding predictions show a Fermi-gas peak at $\pn\lesssim0.25~\gevc$ and the additional Bodek-Ritchie tail~\cite{Bodek:1980ar} (Fig.~\ref{fig:pn}).  As only Fermi motion is considered for such pure QE events, the distribution of $\dalphat$ is flat to first order; a secondary effect arises from the cross section dependence on the center-of-mass energy of the interacting neutrino-neutron system  (Fig.~\ref{fig:dat}).

With IMT, $\dpm$, and therefore $\pn$ [Eq.~(\ref{eq:pn})], is (the modulus of)  the convolution of Fermi motion and IMT [Eq.~(\ref{eq:dpfmimt})]. 
As Fermi motion alone has a small effect on  $\dalphat$, 
the effect of IMT is revealed by the 
 non-flatness of the   $\dalphat$ distribution. 
Following the definition Eq.~(\ref{eq:datdef}) (see also Fig.~\ref{fig:transdia}), it can be seen that when $\left|\ptni\right|<\left|\ptl\right|$, $\dalphat$ becomes larger than $90^\circ$. 
Since in the plane transverse to the neutrino direction the muon and proton  momenta are back-to-back for a free nucleon target (no nuclear effects), $\left|\ptni\right|<\left|\ptl\right|$ means that the proton is (transversely) decelerated by nuclear effects. 
As a result, in Fig.~\ref{fig:dat}, the nominal \genie~prediction with decelerating proton FSI 
 does not contribute greatly to 
the transverse forward boosting region $\dalphat\lesssim90^\circ$~\cite{Lu:2015tcr} where accelerating proton FSI are the dominating IMT. Such accelerating FSI are responsible for the QE peak distortion beyond 5-$\sigma$ total uncertainty at the lowest $\pn$  (Fig.~\ref{fig:pn}).
Of this distortion, the largest contribution comes from the 
 elastic interaction of the \genie~hA FSI model~\cite{Lu:2015tcr} which does not occur in other models, such as \genie's hN FSI model~\cite{Dytman:2011zz}. 
This elastic interaction was designed to resemble hadron-nucleus elastic scattering where the scattered particle could gain energy at the expense of a decelerated recoil nucleus. 
Turning this effect off  can provide a better shape agreement with data.

Non-exclusive IMT---pion FSI and 2p2h---are well separated from the Fermi motion prediction of the $\pn$ QE peak (Fig.~\ref{fig:pn}). At $\pn$ below $0.25~\gevc$,  data points constrain the modeling of Fermi motion; at $\pn \gtrsim 0.4~\gevc$, where pion FSI and 2p2h effects are large, \mnvgenie's prediction follows the data.  
In the transition region, $\pn\sim0.3~\gevc$, \mnvgenie~ shows a clear deficit beyond 1-$\sigma$ total uncertainty. 
With pion absorption and 2p2h events, the measured proton carries a fraction of the total momentum transfer given to multiple particles. Therefore, these reactions 
behave  similarly as decelerating proton FSI   in both $\pn$ and $\dalphat$. The overall \mnvgenie~prediction describes the $\dalphat$ distribution well (Fig.~\ref{fig:dat}).

\begin{figure}[!ht]
\centering
\includegraphics[width=0.494\columnwidth]{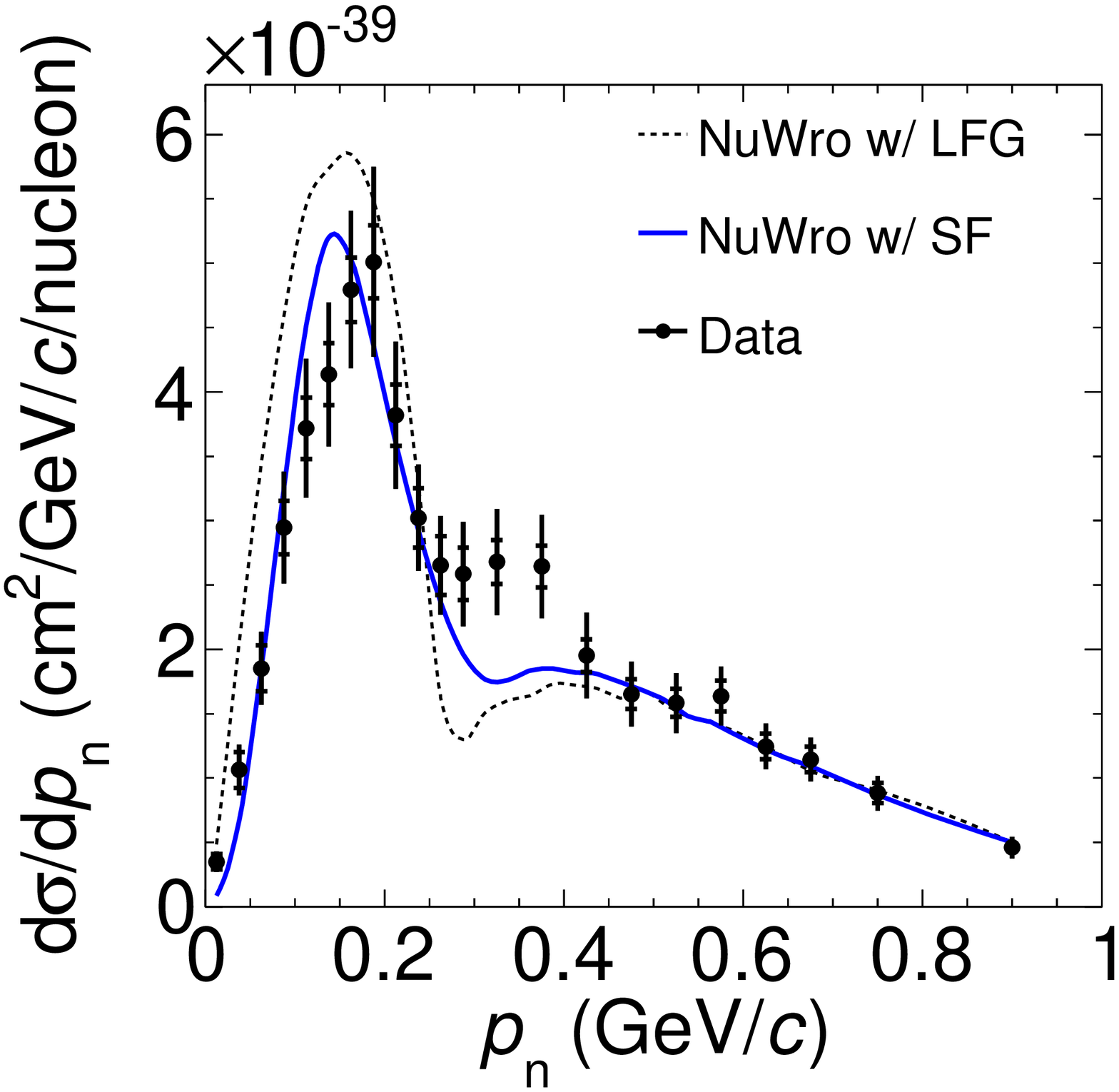}
\includegraphics[width=0.494\columnwidth]{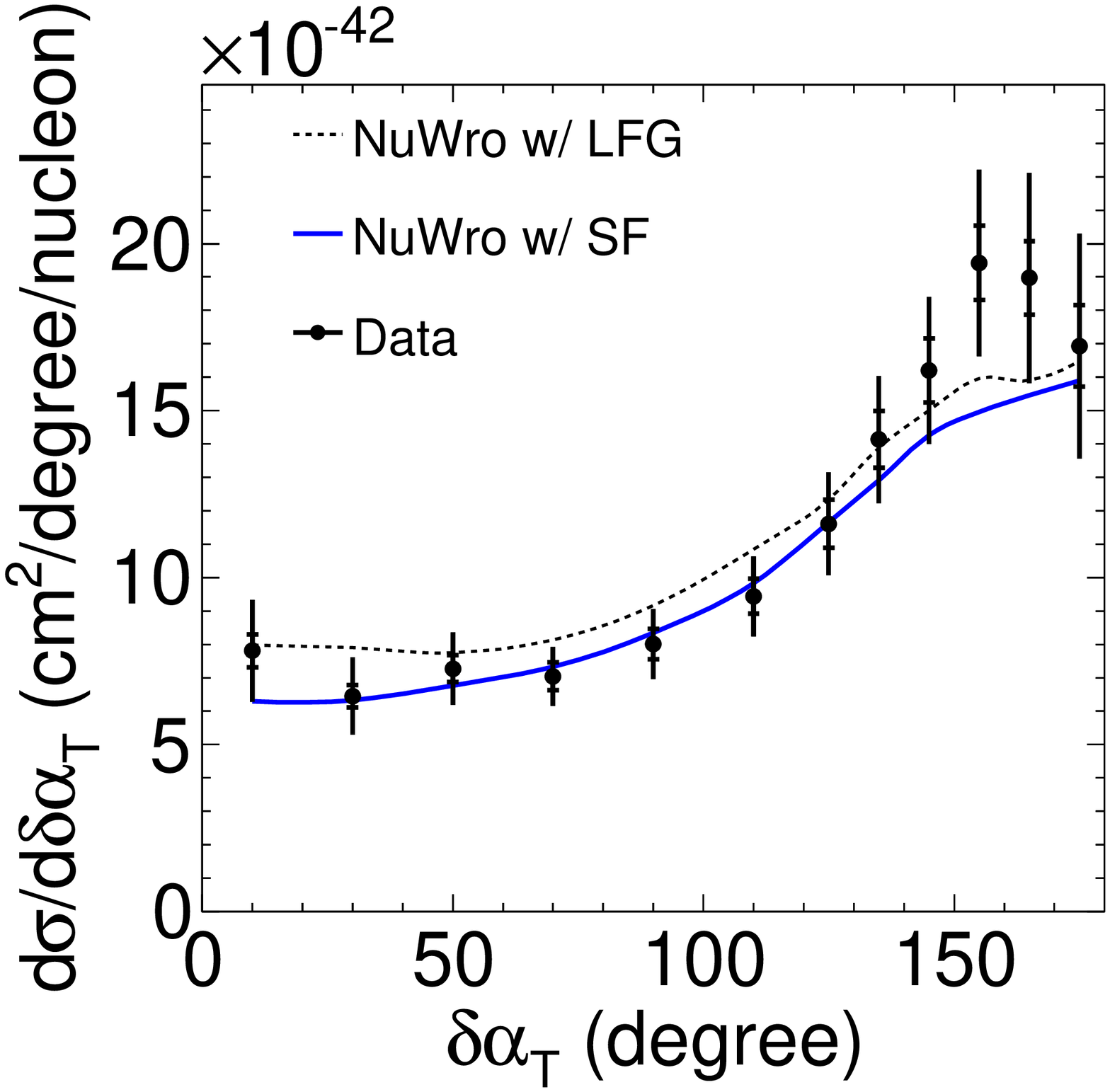}
\caption{Differential cross sections in $\pn$ (\lside) and $\dalphat$ (\rside), compared to \nuwro~predictions with local Fermi gas (LFG) or  Spectral Function (SF).
}\label{fig:nuwro}
\end{figure}

Comparison to the predictions of \nuwro~are shown in Fig.~\ref{fig:nuwro}.  Here, \nuwro's  Spectral Function model works better than its local Fermi gas  model.  Like \mnvgenie, \nuwro~with  Spectral Function is lacking strength at the transition region between Fermi motion and non-exclusive IMT (RES and 2p2h), also with a deficit significantly beyond 1-$\sigma$ total uncertainty.  
  Except for this, the pion production and 2p2h treatment in \nuwro~provide a good description for IMT.  Furthermore, the predictions with  Spectral Function for single particle kinematics are all within 1-$\sigma$ total uncertainties (Supplemental Material~1 Figs.~3 and~4). 

\nuwro's  Spectral Function option does introduce a high-momentum component for the initial nucleon motion, more than the Bodek-Ritchie addition gives to \genie. The  Spectral Function enhancement with respect to  local Fermi gas  appears in the  transition region $\pn\sim0.3~\gevc$.  This is also a kinematic region where the 2p2h tuning process of \mnvgenie~increases the cross section predictions 
up to   $0.4\times10^{-39}~\textrm{cm}^2/\gevc/\textrm{nucleon}$.
 So  introduction of such a tuning process to \nuwro~with  Spectral Function, or a  Spectral Function-based \mnvgenie~might produce a model that agrees better with the data.

In conclusion, the cross sections of QE-like production in terms of $\pn$ and $\dalphat$, whereby the target Fermi motion and the intranuclear momentum transfer are separated, have been presented.  MINERvA's tuned implementation of 2p2h processes gives rate and shape corrections that enable \genie~ to accurately describe the data.  Both \mnvgenie~and \nuwro~with  Spectral Function provide good descriptions of the single particle kinematics and reasonable predictions for $\pn$ and $\dalphat$.  However, both generators fail in the transition region between Fermi motion and IMT dominated regions.  Combination of a \mnvgenie-like 2p2h tune and  Spectral Function might correct this.  

This technique relies on the fine-grained tracking capability of MINERvA to identify and precisely measure ESC protons.  This technique will also be used in experiments with homogeneous trackers such as liquid argon time projection chambers~\cite{Antonello:2015lea, Acciarri:2015uup}. 

Because the variables $\pn$ and $\dalphat$ have sensitivity to initial- and final-state nuclear effects, their measurements on other nuclear targets from MINERvA and from liquid argon experiments should reveal individual A-dependence for Fermi motion and IMT such as FSI and 2p2h, and thus provide a detailed modeling of the nuclear effects that will be valuable for future neutrino oscillation experiments. 

This work was supported by the Fermi National Accelerator Laboratory under US Department of Energy contract No. DE-AC02-07CH11359 which included the MINERvA construction project. Construction support was also granted by the United States National Science Foundation under Award PHY-0619727 and by the University of Rochester. Support for participating scientists was provided by NSF and DOE (USA), by CAPES and CNPq (Brazil), by CoNaCyT (Mexico), by Proyecto Basal FB 0821, CONICYT PIA ACT1413, Fondecyt 3170845 and 11130133 (Chile), by DGI-PUCP and UDI/VRI-IGI-UNI (Peru), by the Latin American Center for Physics (CLAF), by Science and Technology Facilities Council (UK), and by NCN Opus Grant No. 2016/21/B/ST2/01092 (Poland). We thank the MINOS Collaboration for use of its near detector data. We acknowledge the dedicated work of the Fermilab staff responsible for the operation and maintenance of the beam line and detector and the Fermilab Computing Division for support of data processing.

\end{document}